\documentclass[twocolumn,aps,superscriptaddress, showpacs]{revtex4}

\usepackage{graphicx}
\usepackage{dcolumn}
\usepackage{bm}

\newcommand{\pp}{$\pi$ }
\newcommand{\ps}{$\pi + \sigma$ }

\newcommand{\pang}{${\textrm \AA}^{-1}$}
\newcommand{\degree}{$^{\circ}$}

\begin{document}


\title{Effect of Number of Walls on Plasmon Behavior in Carbon Nanotubes}

\author{M. H. Upton}
\affiliation{Condensed Matter Physics and Materials Science Department, Brookhaven National Laboratory, Upton, New York 11973}
\author{R. F. Klie}
\affiliation{Department of Physics, University of Illinois at Chicago, Chicago, IL 60607}
\author{J. P. Hill}
\affiliation{Condensed Matter Physics and Materials Science Department, Brookhaven National Laboratory, Upton, New York 11973}
\author{T. Gog}
\affiliation{%
CMC-XOR, Advanced Photon Source, Argonne National Laboratory, Argonne, Illinois 60439}
\author{D. Casa}%
\affiliation{%
CMC-XOR, Advanced Photon Source, Argonne National Laboratory, Argonne, Illinois 60439}
\author{W. Ku}
\affiliation{Condensed Matter Physics and Materials Science Department, Brookhaven National Laboratory, Upton, New York 11973}
\author{Y. Zhu}
\affiliation{Condensed Matter Physics and Materials Science Department, Brookhaven National Laboratory, Upton, New York 11973}
\author{M. Y. Sfeir}
\affiliation{Condensed Matter Physics and Materials Science Department, Brookhaven National Laboratory, Upton, New York 11973}
\author{J. Misewich}
\affiliation{Condensed Matter Physics and Materials Science Department, Brookhaven National Laboratory, Upton, New York 11973}
\author{G. Eres}
\affiliation{Condensed Matter Sciences Division, Oak Ridge National Laboratory, Oak Ridge, Tennessee  37830}
\author{D. Lowndes}
\affiliation{Condensed Matter Sciences Division, Oak Ridge National Laboratory, Oak Ridge, Tennessee  37830}

\date{\today}

\begin{abstract}
We investigate the physical parameters controlling the low energy screening in 
carbon nanotubes via electron energy loss spectroscopy and inelastic x-ray scattering. 
Two plasmon-like features are observed, one near 9 eV (the so-called 
\pp plasmon) and one near 20 eV (the so-called \ps plasmon).
At large nanotube diameters, the \ps plasmon energies 
are found to depend 
exclusively on the number of
walls and not on the radius or chiral vector.
The observed shift indicates a change in the strength of the
screening and in the 
effective interaction at inter-atomic distances, 
and thus this result suggests a
mechanism for tuning the properties of the nanotube.

\end{abstract}


\maketitle

\section{Introduction}
Carbon nanotubes (CNT) have attracted a great deal of interest lately,
in part because they may be an ideal material for solar cells.
The properties of the exciton are one of the main determinates of 
the efficiency of a solar cell material.
Recent work has demonstrated the profound effect the dielectric 
screening has on the exciton \cite{Perebeinos, Spataru}.
These results are particularly tantalizing because,
contrary to the conclusions of previous 
studies \cite{Stephan, Ajayan, Bursill}, it may be possible
to tune the dielectric function of CNT 
by varying the physical parameters of the nanotubes \cite{Marinopoulos}.

Further, the possibility of tuning the low energy screening in nanotubes
has implications beyond the characteristics of the exciton.
CNTs have a variety of unusual properties, including a 
long electron mean-free-path, 
quantized electronic and phonon bands and possible Tomonaga-Luttinger 
liquid behavior \cite{Saito, Hone, Gao, Tarkiainen, Bockrath}.
Much of this behavior arises because of the near one-dimensionality 
of the nanotubes which in turn increases the importance of 
many-body effects in  the system.
The influence of many-body effects is controlled in part
by the length scale of the low energy interactions.
Thus, the possibility of adjusting these interactions through 
physically accessible parameters is intriguing.

The three key physical parameters affecting  
the electronic properties of nanotubes are
nanotube radius, chiral vector and number of walls:
large curvatures in small-radius nanotube walls
 cause the $\sigma^*$ and $\pi^*$
orbitals to hybridize;
the chiral vector of nanotubes determines whether the
nanotube is metallic or semiconducting; and
finally, the number of walls and the nanotube radius both
provide quantization conditions 
for the electron wave functions \cite{Saito, Sfeir}.
However, it is not well understood what effect, if any,
these parameters have on the length scale of low energy interactions.
The large influence of low energy interactions on the physics
and possible applications of nanotubes
means it is
essential to investigate which parameters control their length-scale
and strength.  



In this paper, the \ps plasmon energies are studied as a function
of nanotube diameter, chirality and number of walls and compared to the 
\ps plasmon energies in bulk graphite.
The plasmon energy reflects the low energy screening
in the sample.
Electron energy loss spectroscopy (EELS) and
inelastic x-ray scattering (IXS) are ideal tools for such an
investigation, as they both directly probe the dielectric
function which determines the screening of
the bare interaction \cite{Handbook}.
By taking advantage of the complementary strengths of 
EELS and IXS
it is possible to
 isolate the effects of 
the three variables and determine that the plasmon frequencies vary 
only with the number of walls.
We show that 
this shift is not an artifact of the increasing intensity
of the surface plasmon with respect to the bulk plasmon, 
as suggested by previous authors
 \cite{Bursill, Ajayan, Stephan, Seepujak}. 
These results imply that the low energy screening of nanotubes may
be tuned simply by changing the number of walls.  

\section{Experiment}
\subsection{Plasmon energy shift with physical parameters}

A large number of 
EELS studies of plasmons in nanotubes have been 
previously performed \cite{Pichler, Kociak, Bursill, Ajayan, Stephan}.  
In particular, the plasmons of single nanotubes \cite{Bursill, Ajayan} 
and of a randomly oriented, mainly single-walled CNT
 mat \cite{Kociak, Pichler} have been studied.
Some of these studies have observed an apparent shift in the \ps plasmon 
energy with increasing nanotube diameter and number of walls.
However, they did not examine the effects of nanotube 
diameter and number of walls separately.


Here we demonstrate that the measured shift
is not a function of the nanotube diameter or chirality 
but the number of walls and
that it is a real shift in the \ps plasmon energy.  
We begin with a discussion of the EELS measurements.

The present EELS measurements 
were performed 
on a number of isolated double-walled
CNTs using an aberration corrected 
scanning transmission electron microscope (JEOL
JEM2200FS), equipped with a Schottky field emission gun operated at
200 keV, an in-column omega energy filter and a CEOS probe corrector. 
The electron probe size (FWHM) was chosen to be 2 $\textrm{\AA}$ to
provide sufficient electron beam intensity to study single nanotubes.  
The collection and convergence angles were 8.9 and 15 mrad respectively.
The energy resolution was measured to be 1.0 eV. 
The beam spot was placed on the center of the tube
so that the momentum transfer, which is perpendicular to the beam direction, 
was tangential to the walls of the nanotube. 

\begin{figure}
\includegraphics[width=3in]{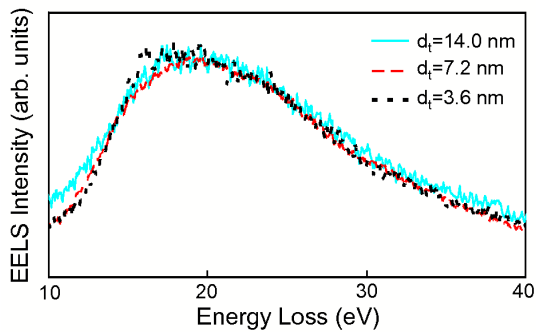}
\caption{\label{fig:EELS} (color online)
EELS spectra from three double walled carbon nanotubes of different diameters.
The blue, solid line 
corresponds to a diameter of 14.0 nm;
 the black, short-dashed line to
a diameter of 7.2 nm and the red, long-dashed line to 
a diameter of 3.6 nm.
The spectra are all taken at the same position in the Brillouin zone and
have been normalized to the sample peak intensities. The elastic line 
has been subtracted off in each case.
}    
\end{figure}

The resulting energy loss spectra are shown in figure \ref{fig:EELS}.
These data in fact represent an
integral over a disk 
in reciprocal space with a radius of $Q=5.5{\textrm \AA}^{-1}$ 
because of the large acceptance of the detector and the beam convergence angle.
Interpreting the
spectra quantitatively is further complicated by
multiple scattering 
effects, though based
on a comparison of the measured plasmon energies with
the plasmon energy dispersions measured with IXS, discussed below,
the EELS 
intensity of figure 
\ref{fig:EELS} appears to be dominated by contributions from close to $Q=0$.
For the purposes of the present study,
however, the
important point is that all the EELS spectra are measured 
under identical conditions
and therefore may be directly compared with each other. 

We now discuss
effect of the diameter and chirality of the CNT
(which changes the curvature of the walls) on the \ps plasmon energies. 
Figure \ref{fig:EELS} shows measurements from three different, 
isolated, double-walled CNT
with diameters of $3.6$, $7.2$ and $14.0$ nm.
These tubes were taken from the same aligned nanotube arrays 
discussed below.  
The \ps plasmon is clearly visible near 18 eV.  
Despite the drastically different diameters of the 
nanotubes, the \ps plasmon energies are identical.
Although the chirality of the nanotubes was not explicitly checked,
it is impossible for nanotubes with different diameters to have the same chiral
vectors.  Further, it is less than 1\% probable that even the two
 smallest nanotubes would have the same chiral angle.
These results therefore explicitly demonstrate, for the first time, that the plasmon energy is 
independent of nanotube diameter and chirality, for
large enough diameters.

\begin{figure}
\includegraphics[width=3 in]{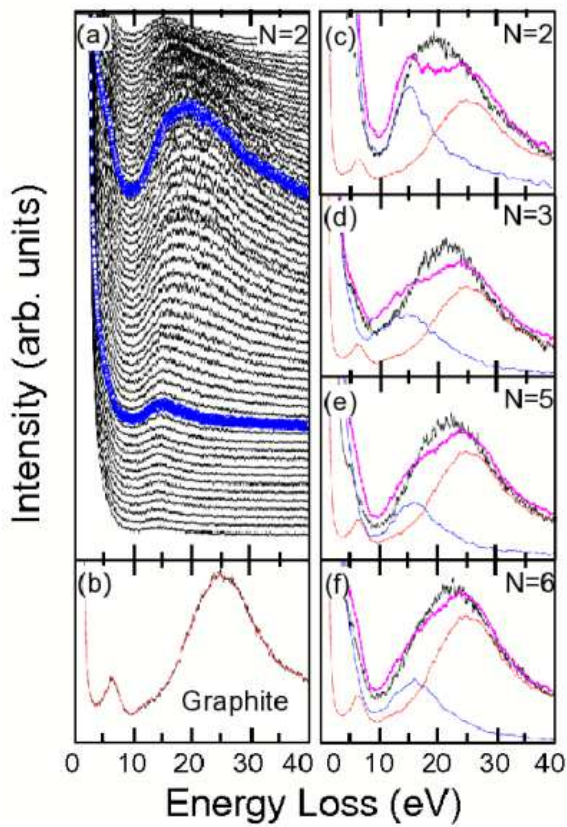}
\caption{\label{fig:Surface} (color online)
(a) EELS spectra from a double-walled nanotube. 
 Different spectra are from different points across the tube. 
Spectra corresponding to the edge and center of the tube
are displayed in blue circles.
(b) Raw EELS spectrum from graphite.
(c) The center spectrum from a double-walled nanotube (black).
The smoothed edge
spectrum from this tube
is shown in blue and the smoothed graphite spectrum is shown in red.
The sum of the edge spectrum and smoothed graphite is shown in purple.
(d), (e) and (f) The same as (c) for 3, 5 and 6 walled nanotubes.
}    
\end{figure}

In figure \ref{fig:Surface} a series of EELS scans taken on 
individual nanotubes with varying number of walls is shown.
The \ps plasmon energy seems to be dependent on the number of walls of a CNT.
The black line in figure \ref{fig:Surface}c-f shows EELS data from
 the center of nanotubes with 2, 3, 5 and 6 walls respectively.
Clearly, the plasmon peak energy shifts with the number of walls.
The apparent shift in the \ps plasmon energy is therefore exclusively
a function of the number of walls and not of the nanotube diameter
or chiral vector.

By taking spatially
resolved EELS measurements we have been able to show that
the energy shift in the \ps plasmon is an actual
shift in the plasmon energy and not an artifact of the increased 
amplitude of the surface plasmon relative to a constant energy \ps plasmon
state as the number of walls decreases, as suggested in previous 
studies \cite{Bursill, Stephan, Ajayan}.
Figure \ref{fig:Surface}a shows a series of EELS spectra taken at different 
positions across a
double-walled nanotube.  This data is qualitatively similar
to data in \cite{Stephan}.  
The spectra from the edge of the tube and the center 
of the tube are displayed in blue circles. 
The peak in the edge spectrum
has been identified as the surface plasmon state in previous
publications \cite{Bursill, Stephan}.  
Figure \ref{fig:Surface}b shows an EELS spectrum of graphite.
Figure \ref{fig:Surface}c shows the center spectrum from the double-walled
nanotube in black, a smoothed graphite spectrum (red), a smoothed
edge spectrum (blue) and the sum of the graphite and edge spectrum (purple).
It is apparent that the \ps plasmon peak in the double-walled nanotube is not
a combination of the edge state and the graphite \ps plasmon.  
Figures \ref{fig:Surface}d-f demonstrate this point for 
nanotubes with 3, 5 and 6 walls.  Even for nanotubes with 6 walls, 
the addition of the graphite and edge spectra
give a peak with an energy several eV higher than the measured spectrum.
Thus, these EELS spectra show that for a 
sufficiently low number of walls 
the measured \ps plasmon peak can not be the sum 
of a surface plasmon peak and a graphite-like \ps plasmon peak.

\subsection{Momentum dependence}
Having established the existence of a shift of the \ps plasmon energy with the
number of walls we turn to the momentum dependence of the plasmon shift,
and thus the momentum dependence of the change in screening.
This momentum dependence 
can not be measured with EELS.
Although, as illustrated above, one of the great strengths of EELS  is 
its ability to study
one nanotube at a time, it
suffers from limited momentum range due to 
multiple scattering effects which complicate the data analysis
at large momentum transfers.
Additionally, in previous studies, orientation information
has often been limited by
the sample alignment and geometry.
To address the momentum dependence of these features therefore,
IXS and aligned nanotube samples are utilized.

IXS, like EELS, 
measures the dynamical structure factor, $S(Q,\omega)$.  
The advantage of IXS is
that
the small x-ray scattering cross-section means that there are almost no 
multiple scattering inelastic events and 
the interpretation of the data
is straightforward, even at high momentum transfers.
However, IXS cannot look at individual nanotubes but
requires an ensemble of nanotubes.  
The technique is described in detail elsewhere \cite{Schulke, Handbook}.
The samples studied here were 
arrays of long, aligned carbon nanotubes \cite{Eres}.
These samples 
allow for a significant advance over earlier 
ensemble measurements because
they permit measurements of nanotubes in one
 orientation, rather than a range of orientations.
In these experiments, the resolution of the absolute value of $Q$ is 
set by the intrinsic resolution of the instrument, 0.056 \pang.
However, the imperfect alignment of the carbon
nanotubes gives rise to uncertainty in the direction of $Q$.
Nevertheless, the combination of IXS and 
aligned samples allow the excitations to be measured with a larger range of
reciprocal space than has been previously possible.
We note that this is the first IXS study of carbon nanotubes, 
though there have been previous IXS studies of the plasmons of graphite \cite{Schulke, Hiraoka}.

\begin{figure}
\includegraphics[width=3in]{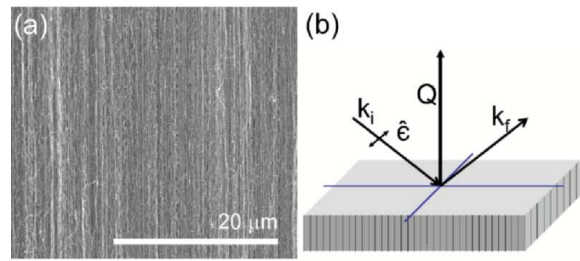}
\caption{\label{fig:geometry}
(color online) (a) Scanning electron microscope picture of an aligned 
array of multi-walled carbon nanotubes  and (b) schematic of the  
experimental geometry for the x-ray experiments.  
The axes of the nanotubes are parallel 
to the direction of momentum transfer, $Q$.}
\end{figure}


 The aligned CNT samples studied here were
grown 
on a Si(001) substrate by chemical vapor
 deposition \cite{Eres}.  
Two types of samples were studied: aligned multi-walled and aligned few-walled
carbon nanotube (MW- and FW-CNT) arrays.
The tube axes in each case are
 within 20 degrees of the substrate normal.  A scanning electron microprobe
 picture of one such MWCNT sample is shown in figure \ref{fig:geometry}a.  
The nanotubes have
 a relatively low packing fraction ($10 \textrm{\%}-17 \textrm{\%}$)
 and are not bunched in ropes, as
 verified by
 x-ray transmission and microprobe measurements.  Therefore, because
the nanotubes are too separated to strongly interact, 
 the results reported here are representative 
of individual nanotubes and not of a bundle
of interacting nanotubes.  
This is an important point because 
the environment of a nanotube, for example, a substrate that a nanotube 
rests on, can change the properties of a nanotube \cite{LeRoy, Marinopoulos}.

The physical properties of these nanotube arrays have
been characterized 
by a number of techniques including TEM and Raman spectroscopy.
In particular, a series of TEM images show that 
the MWCNT sample comprises tubes with an average outer diameter of 
$19 \pm 6 $ nm and an average of $14 \pm 5$ walls,
while the nanotubes in the FWCNT sample have an average outer 
diameter of $5 \pm 2$ nm and $3 \pm 2$ walls. 
These parameters are consistent with Raman experiments, which
measured breathing modes, charactieristic 
of few-walled NT, in the FWCNT sample but not in the MWCNT sample.

The IXS experiments reported here were performed at beam line 9-ID,
CMC-XOR,  at the Advanced Photon Source.  
The overall resolution of the spectrometer was 300 meV (FWHM).
The incident photons were polarized, perpendicular 
to the scattering plane. 
The polarization was thus perpendicular to the axis of the nanotubes 
(figure \ref{fig:geometry}b).
Energy scans were performed by varying the incident energy while holding 
the final energy fixed at 8.9805 keV.
In all the data shown here, the direction of the 
momentum transfer was along the axes of the aligned nanotube samples.
Spectra were taken at room temperature.

\begin{figure}
\includegraphics[width=3in, height=3in]{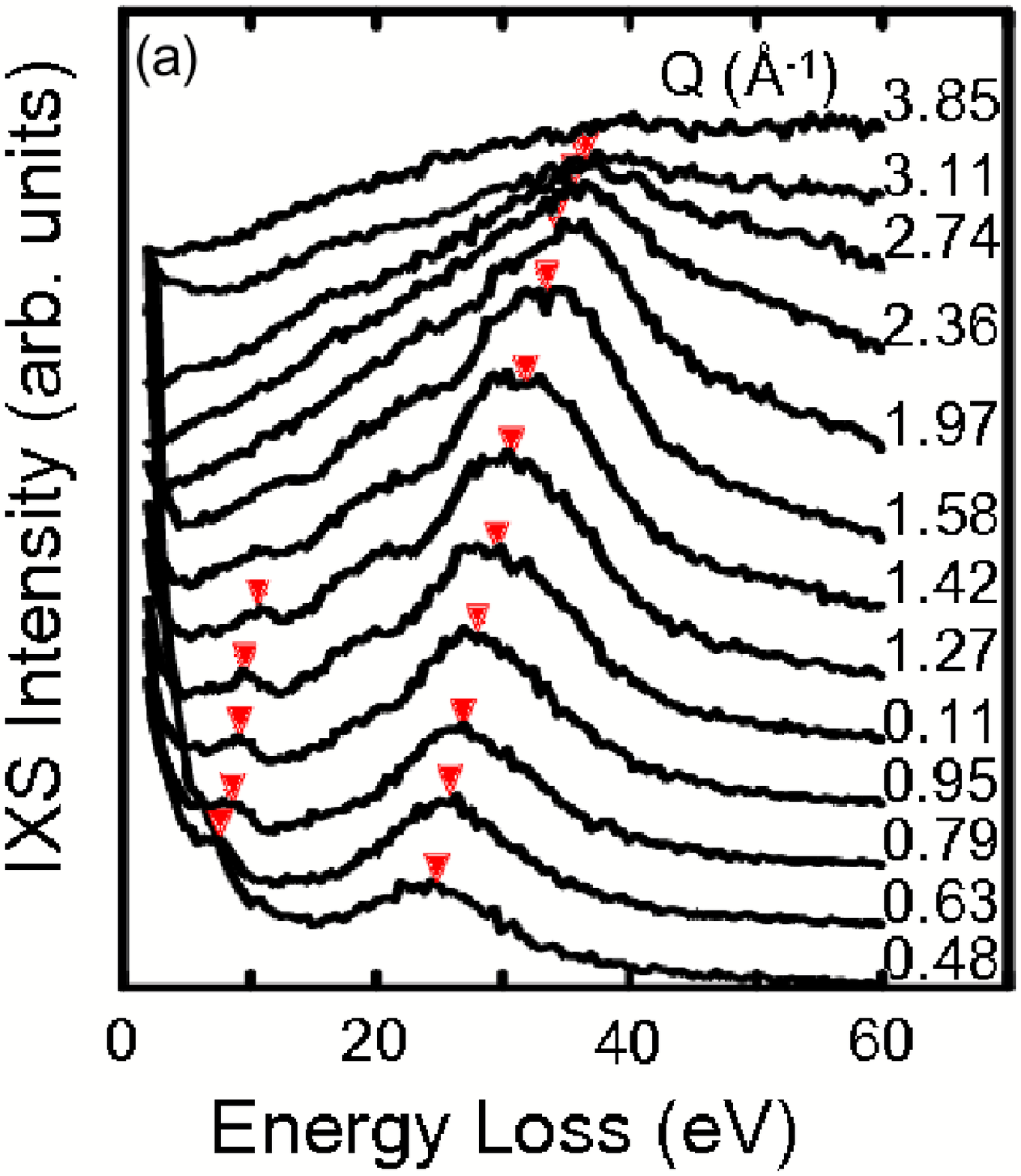}
\includegraphics[width=3in, height=3in]{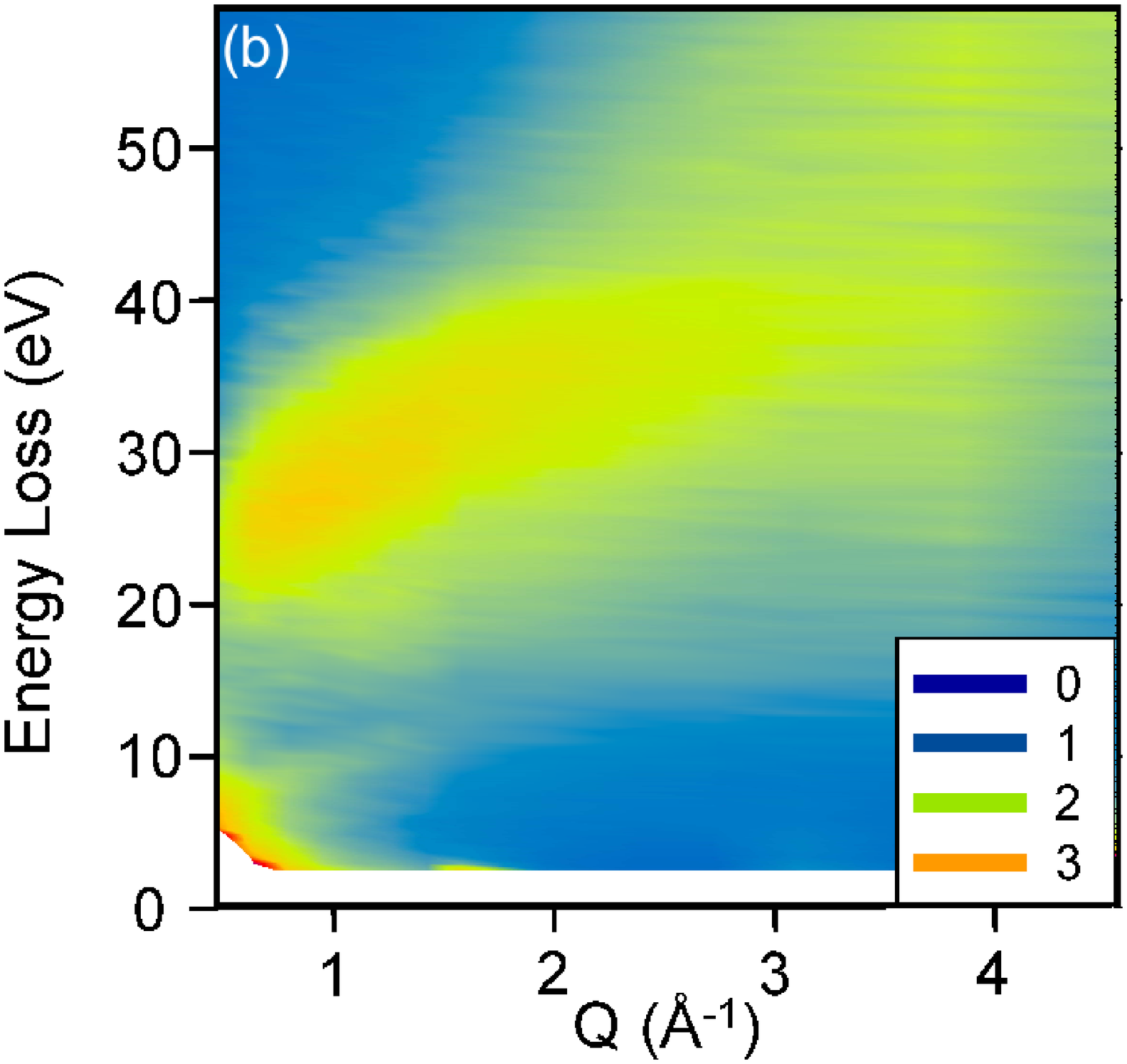}
\caption{\label{fig:spectra} (color online)
(a) Inelastic x-ray scattering spectra from the few-walled carbon nanotube 
sample.
  $Q$ is parallel to the nanotubes' axis.  
The \ps plasmon is the large feature between 20 and 40 eV energy loss.  
The \pp plasmon is the small peak near 10 eV energy loss.  
The elastic lines have been omitted in these plots.  
A red triangle points  to the peak positions.
Data are offset vertically for clarity.
(b) The same data as a contour plot.
}
\end{figure}

Representative IXS spectra of the FWCNT sample are 
shown in figure \ref{fig:spectra} \cite{AngleNote}.  
Both the \ps ($\sim 20$ eV) and \pp ($\sim 9$ eV) 
plasmon bands are visible.
To extract the plasmon peak positions the spectra were fit with 
Lorentzians.  
The quality of the \ps fits to some of the data taken 
at intermediate momentum transfers was improved by 
fitting a low energy loss shoulder near 22 eV and a small amount of 
inter-band structure near 28 eV with additional Lorentzians \cite{Hiraoka}. 
The addition of these small Lorentzians does not affect 
the results of this paper, which focuses on the
 behavior of the large \ps plasmon.


We note in passing that carbon nanotube properties are known to
 be sensitive to the 
surface adsorption of water and other atmospheric gases 
\cite{Zahab, Collins, Kong}.  
Surface adsorption effects were therefore looked for in the IXS data
by outgassing 
a sample in rough vacuum.
Specifically, a sample was heated to 120\degree C over 90 minutes
 and then held at 120\degree C 
for 3 hours.  There was no measurable difference between the spectra
before and after this outgassing procedure.


\begin{figure}
\includegraphics[width=3.in]{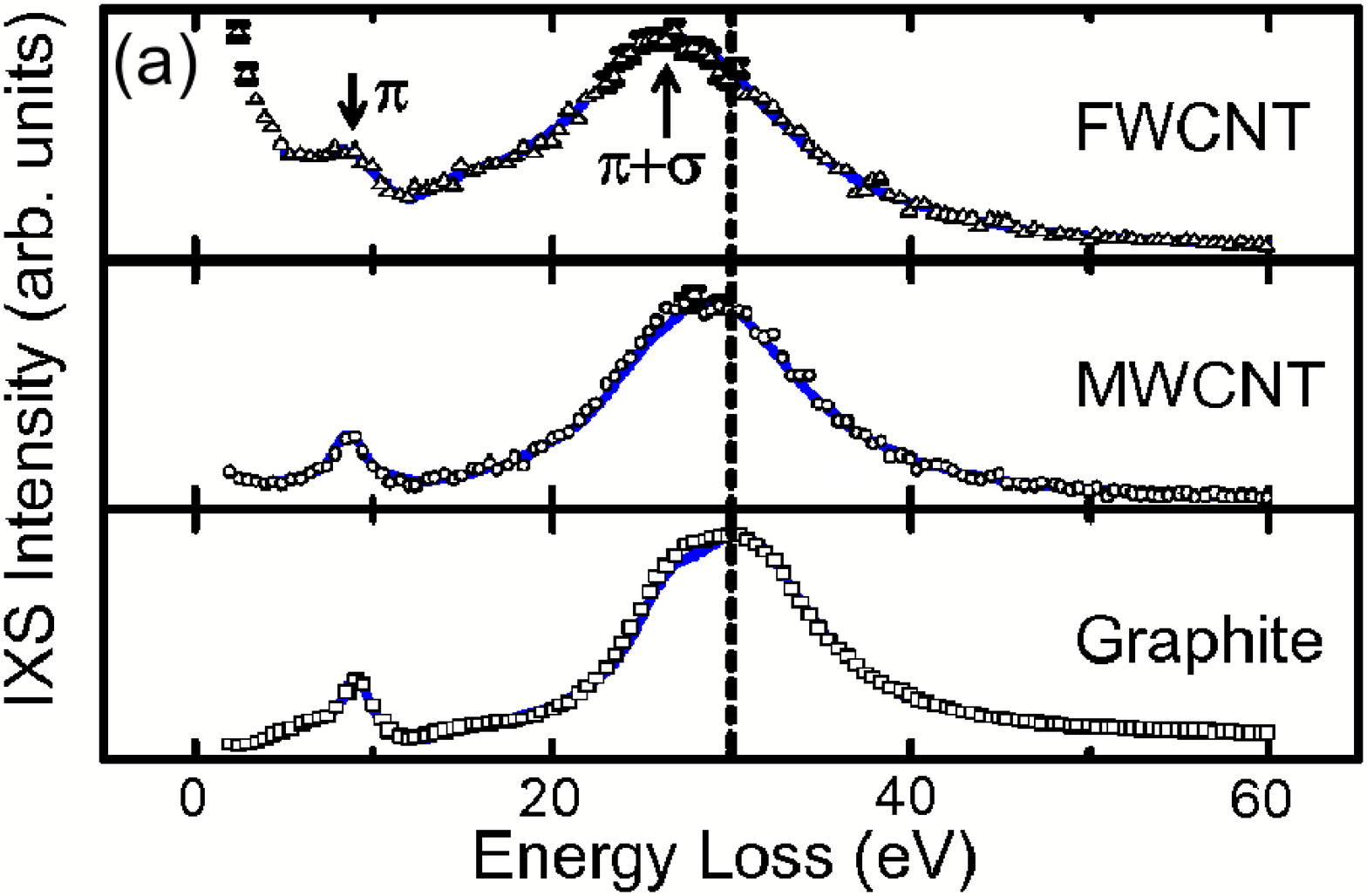}
\includegraphics[width=3.in]{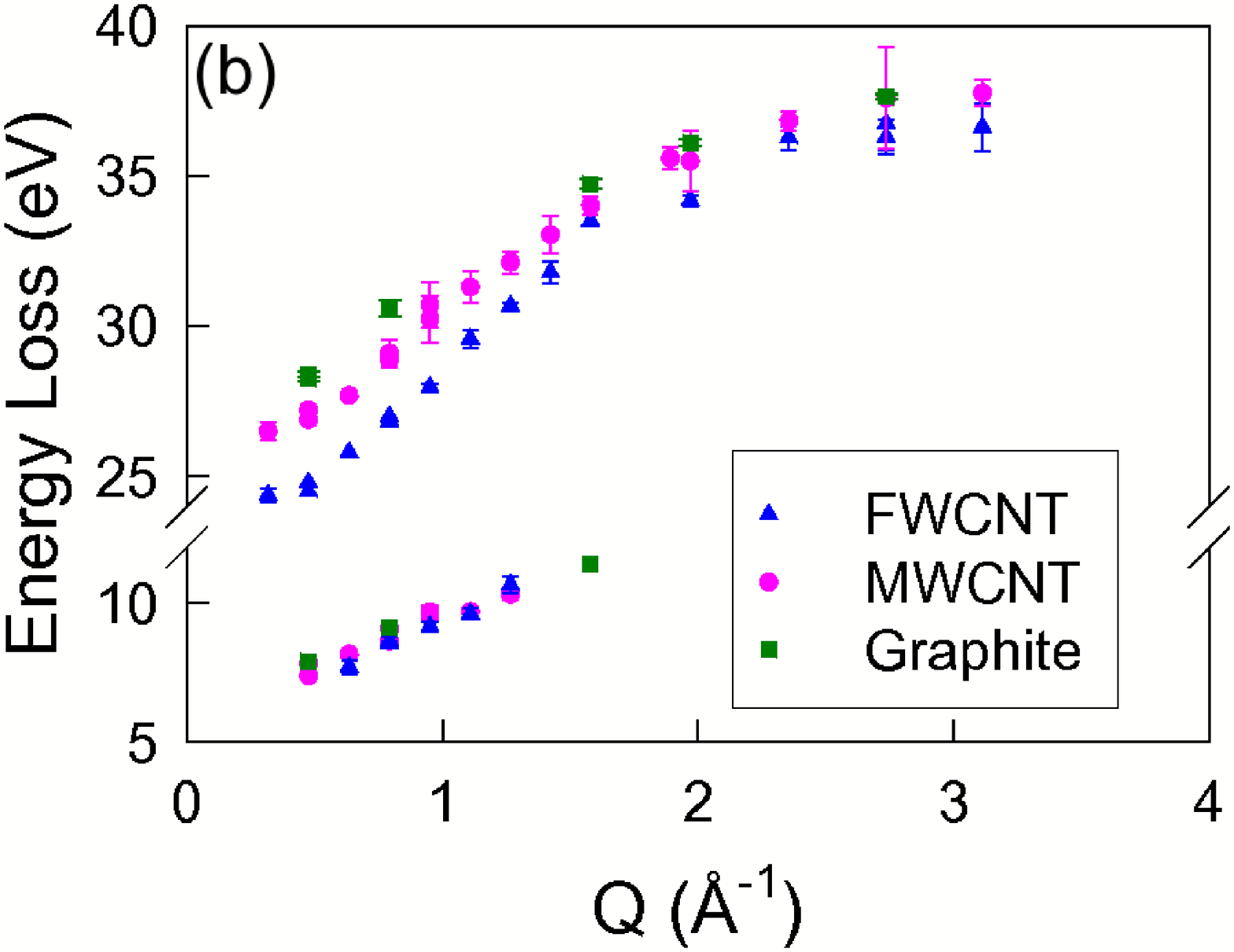}
\caption{\label{fig:disp} (color online)
(a) Inelastic x-ray scattering spectra 
from few-walled (top panel), multi-walled carbon nanotube (middle panel) 
and HOPG (bottom panel).
In each case, $Q=0.79$\pang with $Q$ along the nanotubes' axis and in
the plane of graphite, respectively.  
The elastic lines have been omitted in this plot.
Note that the tail of the elastic line in the 
FWCNT spectrum appears unusually large simply because
there is less FWCNT material and the inelastic signal 
is therefore correspondingly smaller than in other cases.
(b) Plasmon dispersion for the three samples.  
The error bars come from the Marquardt least-squares fitting algorithm.  
When no error bars appear, the error is smaller than the symbol size.
}    
\end{figure}

In figure \ref{fig:disp}a, representative IXS spectra for the
FWCNT, MWCNT and HOPG  samples are shown.
In each case, the data were taken at 
$Q = 0.79 $\pang.  For the nanotube samples, the momentum transfer 
was along the axes of the nanotubes and for
 the graphite sample the momentum transfer was in the plane of 
the graphite sheets.  
As was the case in the EELS data, the \ps plasmon is observed to
shift with the number of walls.


\section{Discussion}

The most obvious explanation of 
the \ps plasmon shift is that the electron
density in the nanotubes decreases as the number of walls decreases.
The simplest model of plasmons, the jellium model, predicts
the plasmon energy, $\omega_p$, at the center of the Brilluoin
zone varies with the electron density, $n$, as
$\omega_p \sim \sqrt{n}$ \cite{Marder}.
To explain the observed \ps plasmon
shift within the context of the jellium model
 the FWCNT sample would have to have only 76\% 
of the electron density of graphite, an exceptionally
 large decrease.

A lower electron density may result from a number of causes. 
The decreased density of graphite 
in carbon nanotubes
may reduce the electron density appreciably.  The
graphene-graphene distance is larger in nanotubes than in graphite and
increases at very small nanotube diameters \cite{Kiang}.  
However, the predicted
difference between the FWCNT and MWCNT densities
is less than 1\% and the  difference between the MWCNT and graphite
densities is less than 3\%.

Another possible cause of the decreased electron density is that
the surface tails of the electron wave functions
have significantly reduced the electron density in the bulk.  
When there are only a few
walls the surfaces have a greater influence on the nanotube
properties than when there are effectively an infinite number of walls.  
However, to get any reasonable agreement
with the data the electron tails of the surface layers
need to be excessively large:
the integrated intensity of the surface tails of a surface sheet
would be
 half of the magnitude of the integrated electron intensity of one
interior sheet.

Perfect quantitative agreement
with a simple model like the jellium model should not
be expected.  In this case, however, the disagreement is extremely 
large and one must look for a different explanation.


Instead, we
propose that the plasmon energy shift is driven by
changes in screening.
The momentum dependence of the plasmon energy for the three cases is shown
in figure \ref{fig:disp}b.
At large $Q$, corresponding to small length scales, the difference
between graphite, MWCNT and FWCNT \ps plasmon energies 
is small or nonexistent.  This makes sense.
At these (atomic) length scales, screening
effects would be expected to be small, and equal in the three samples.
At small and intermediate $Q$, corresponding to longer length scales, 
the differences in the plasmon energies are significant,
with samples with fewer walls exhibiting a significantly lower plasmon energy
than those with more walls.  The
energy difference becomes noticeable for $Q\leq2$ \pang, which corresponds
to length scales greater than $3$  $\textrm{\AA}$, 
near the interplanar
 distance in graphite. 
As demonstrated above, the \ps plasmon energy depends only on the
number of walls.
The momentum dependence shows this effect is only present for length
scales greater than the interplanar spacing.  
This again makes intuitive sense.
At shorter length 
scales, the presence or absence of neighboring sheets of graphite 
is not felt, and plasmon energies converge.

The plasmon energy is determined by the zero of the real 
part of the energy and
wave-vector dependent dielectric function, $\epsilon(\bf{Q},\omega)$.
From a general understanding of the 
dielectric constant, 
our result
implies that, at low energies,
the magnitude of the 
dielectric function is smaller in samples with fewer walls.
This indicates that the length scale of low energy interactions is longer
in FWCNT than in MWCNT, and longer in MWCNT than in graphite.
There are a number of consequences of changes
in this length scale for the
optical properties of the system.
For example, it implies that excitons in MWCNT are more likely to dissociate
than excitons in FWCNT, because the additional screening in MWCNT decreases 
the binding energy of excitons.



\section{Summary}
We have shown that for diameters larger 
than 3.6 nm, the \ps plasmon energies near $Q=0$ of CNT samples depend 
exclusively on the number of walls, and that the shift observed is
not an artifact
of a relative change in intensity of the surface plasmon.  
That is, a unique \ps plasmon dispersion 
and a unique $Q=0$ plasmon energy 
 are associated with a particular number of walls.
We infer that the low energy screening, which controls the physical
extent of the low energy interactions,
is tuned by changing the number of walls in carbon nanotubes.

\section{Acknowledgments}
The authors thank C.~C.~Homes and L.~Carr for optical characterization 
of the samples, A.~Stein for electron microprobe pictures of the samples and 
J.~Fink for helpful discussions.
Work performed at BNL was supported by US DOE, Division of
Materials Science and Engineering, under contract No.~DE-AC02-98CH10886
and partially by DOE-CMSN.
Use of the Advanced Photon Source was supported by the US DOE, 
Office of Science, Office of Basic 
Energy Sciences, under Contract 
No.~W-31-109-Eng-38.
Research performed at ORNL was sponsored by the Division of 
Materials Sciences and Engineering,
 Office of Basic Energy Sciences, US DOE, under contract
 DE-AC05-00OR22725.

\bibliography{Upton}

\begin{thebibliography}{26}
\bibitem[{Ang()}]{AngleNote}
\bibinfo{note}{The spectra are normalized by incident x-ray intensity and a
  geometric factor, $1/\cos{\theta}$, which accounts for variations in the beam
  footprint on the sample as the scattering angle is varied.}
\end{thebibliography}
\end{document}